\documentclass{article}
\usepackage{graphicx,color}% Include figure files
\usepackage{dcolumn}% Align table columns on decimal point
\usepackage{bm}% bold math
\usepackage{color,comment,amsmath,amssymb}
\newcommand\ds{\displaystyle} % differential
\newcommand\R{{\bf R}}
\newcommand\M{{\bf M}}
\newcommand\Me{{\bf M_e}}
\newcommand\W{{\bf W}}
\newcommand\A{{\bf A}}
\renewcommand\S{{\bf S}}
\begin{document}
\title{Wall influence on dynamics of a microbubble}

\author{Sergey A.~Suslov\footnote{Mathematics, H38, Swinburne
    University of Technology, Hawthorn, Victoria 3122, Australia},
  Andrew Ooi\footnote{Department of Mechanical Engineering, University
    of Melbourne, Melbourne, Victoria, 3010, Australia}, 
  Richard Manasseh\footnote{Mechanical Engineering, H38,
    Swinburne University of Technology, Hawthorn, Victoria 3122, Australia}}
\date{ }
\maketitle

\begin{abstract}
  The nonlinear dynamic behaviour of microscopic bubbles near a wall
  is investigated. The Keller-Miksis-Parlitz equation is adopted, but
  modified to account for  
  the presence of the wall. This base model describes the time
  evolution of the bubble surface, which is assumed to remain
  spherical, and accounts for the effect of acoustic radiation losses
  owing to liquid compressibility in the momentum conservation. 
  Two situations are considered: the base case of an isolated bubble
  in an unbounded medium; and a bubble near a solid wall. In the latter
  case, the wall influence is modeled by including a symmetrically
  oscillating image bubble. The bubble dynamics is traced using a
  numerical solution of the model equation. Subsequently, Floquet
  theory is used to accurately detect the bifurcation point where
  bubble oscillations stop following the driving ultrasound frequency
  and undergo period-changing bifurcations. Of particular interest is
  the detection of the subcritical period tripling and quadrupling
  transition.  The parametric bifurcation maps are obtained as
  functions of non-dimensional parameters representing the bubble
  radius, the frequency and pressure amplitude of the driving
  ultrasound field and the distance from the wall. It is shown that
  the presence of the wall generally stabilises the bubble dynamics, so that
  much larger values of the pressure amplitude are needed to generate 
  nonlinear responses.
\end{abstract}

\section{Introduction}
When driven by the oscillating pressure field of ultrasound,
microbubbles strongly scatter the incident acoustic waves, and can
resonate or fragment \cite{deJong2002,Brennen2002}. Intraveneously injected
microbubbles have been used in clinical practice for over two decades
\cite{Lindner2004} because their physical response makes the blood that
transports them stand out in the scan relative to the surrounding
tissue: this is ``contrast'' to the clinician. Modern microbubble
contrast agents are coated in an elastic shell that retards its
dissolution \cite{Grinstaff1991,Stride2008} and also affects its
properties \cite{deJong2002,Postema2007}.

Targeted ultrasound contrast agents are microbubbles in which the shells are
coated with molecules, usually antibodies, that adhere to specific disease
markers \cite{Klibanov2009}. So far, they are not in clinical use. Despite
significant research into the nature of the signal when microbubbles are bound
to targets \cite{Doinikov2009,Qin2009,Lindner2009}, at present, there is no
clinically-accepted way to tell adherent microbubbles from free microbubbles in
real time \cite{Zhao2006,Patil2009,Patil2011}. Since the microbubbles
have a short lifetime, rapid discrimination of those that are attached
to target walls from those that have not would be an important step
towards clinical practice \cite{Lindner2009}. Suggestions have
included filtering based on the fact that the speed of adherent
microbubbles should be zero \cite{Patil2009,Patil2011}, although this
does not discriminate between microbubbles in the blood-vessel
boundary layer and ones truly adhered. Alternatively, the acoustic
radiation force could be used to ``push'' microbubbles; only the free
microbubbles would move, enabling the difference from bound
microbubbles to be discerned \cite{Zhao2007}. 

\begin{figure}
  \centerline{\includegraphics[width=0.4\textwidth]{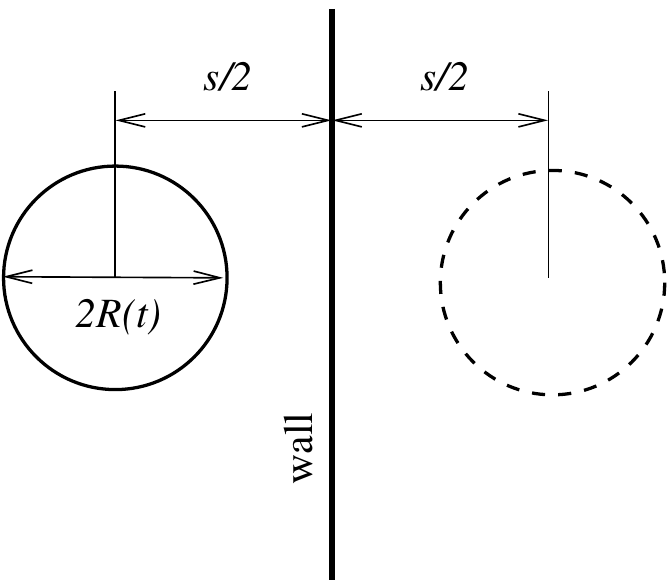}}
  \caption{Schematic view of the problem geometry.\label{geom}}
\end{figure}

For real-time detection of adherent agents, a further suggestion is to exploit
alterations in the linear response frequency \cite{Payne2011} that characterise
adherent bubbles. If viscous forces acting in the fluid near the wall are
neglected, potential-flow theory allows the wall to be replaced with an
identical ``mirror'' bubble image that is located symmetrically with respect to
the wall plane, see Fig.~\ref{geom}, and oscillates with the same frequency,
amplitude and phase as the original bubble \cite{Manasseh2009}. This
mirror-pair oscillates in the symmetric coupled-bubble mode
\cite{Strasberg1953,Payne2005}. If the variable measured is the sound
pressure in the liquid, the linear natural frequency of the bubble is shifted
downwards by a factor of $\sqrt{2/3}\approx0.82$ \cite{Manasseh2009}. In
the medical application that motivated the present work, the walls of
blood vessels are unlikely to be rigid; in fact the mirror-image
symmetric mode for a rigid-wall case is one extreme. The opposite
extreme is when there is a free surface when the bubble and its image
oscillate in  the anti-symmetric mode
\cite{Manasseh2009,Pumphrey1990,Illesinghe2009}. 

Recent experiments with high-speed imaging demonstrated a downshift in the
linear response frequency of actual targeted contrast agents on biological
target surfaces \cite{Overvelde2011}, broadly consistent with suggestions based on
theory and experiments on microbubbles larger than contrast-agent sized
\cite{Payne2011}. With current microbubbles, the diagnostic specificity of such
techniques would be poor, because the microbubbles vary greatly in size and
thus resonant freqency; and monodispersed microbubbles are still experimental
constructs \cite{Dollet2008,Chen2009}. Thus, a suite of further indicators of
proximity to target walls would be beneficial. Adding further indicators
to the techniques already suggested in the literature should improve the 
sensitivity and specificity of microbubble targeting.

Detailed research into the nonlinear dynamics of a single, isolated
microbubble began with the work of Lauterborn in 1976
\cite{Lauterborn1976} and by 1990 detailed bifurcation diagrams were
calculated illustrating the complex behaviour possible as driving
frequency and amplitude were changed \cite{Parlitz1990}. Recently, the
effect of acoustic coupling with neighbouring bubbles was examined
numerically in \cite{Chong2010} using a coupled-oscillator approach.
In the present paper, we expand on the earlier suggestions for detection of 
microbubbles on walls based on linear theory, to a study of the
nonlinear dynamics of a microbubble in the vicinity of a solid
wall. In particular, we discuss the non-uniqueness of nonlinear bubble
oscillations that occur for the same driving parameters but different
initial conditions.

In addition to the neglect of viscosity inherent in the mirror-image theory, we
also assume that the distance $s/2$ between the wall and the bubble centre
remains constant (in reality, Bjerknes forces would cause this distance to vary
\cite{Mettin1997,Vos2008}). The effects of the bubble shell, which differ widely
amongst microbubble types (and differ amongst models of the shell) 
\cite{Allen2002,Paul2010}, are also neglected. Such assumptions are
made to simplify the calculations and may be refined in the
future. For the present, the immediate aim is simply to determine if
potentially useful differences exist in the nonlinear dynamics between
adherent and free microubbles. As noted earlier, the ultimate aim is 
to derive further criteria of wall proximity.

\section{Modified Keller-Miksis-Parlitz equation}
%We study the dynamics of a microbubble in the vicinity of a
%solid wall. As a first approximation we neglect viscous forces
%acting in the fluid near the wall. Under this assumption the wall
%is replaced with an identical bubble image that is located
%symmetrically with respect to the wall plane, see Fig.~\ref{geom},
%and oscillates with the same frequency, amplitude and phase as the
%original bubble. We also assume that the distance $s/2$ between the
%wall and the bubble centre remains constant.
The base model employed in the current study is Keller-Miksis-Parlitz
\cite{Keller1980,Parlitz1990} equation modified to account for the
presence of an image bubble, see the last term in the right-hand side
of equation (\ref{KMPE}), 
\begin{eqnarray}
  &\left(1-\frac{\dot{R}}{c}\right)R\ddot{R}
  +\frac{\dot{R}^2}{2}\left(3 -\frac{\dot{R}}{c}\right)&\nonumber\\
  &=\frac{1}{\rho}\left[1+\frac{\dot{R}}{c}
    +\frac{R}{c}\frac{d}{dt}\right]\left[P\left(R,\dot{R}\right)
    -P_\infty(t)\right]-{\frac{1}{s}\left(R^2\ddot{R}+2R\dot{R}^2\right)}\,,&
  \label{KMPE}
\end{eqnarray}
where
\begin{equation}
  P\left(R,\dot{R}\right)
  =\left(P_0-P_v+\frac{2\sigma}{R_0}\right)\left(\frac{R_0}{R}\right)^{3\kappa}
  -\frac{4\mu\dot{R}}{R}-\frac{2\sigma}{R}\,.
\end{equation}
The expression $P_\infty(t)=P_{0}-P_v+\alpha \sin(\omega t)$, where
$\omega=2\pi f_e$, represents the pressure in the liquid far from the
bubble, and $R(t)$, $R_0$, $\mu$, $\rho$, $\kappa$, $c$, $\sigma$,
$\alpha$ and $f_e$ denote the instantaneous bubble radius, the
equilibrium bubble radius, the dynamic viscosity of the liquid, the
density of liquid, the polytropic exponent of a gas entrapped in the
bubble, the speed of sound in the liquid, the surface tension of a gas/liquid
interface, the acoustic pressure amplitude and the driving frequency,
respectively. The model considered accounts for the decay of
  bubble oscillations due to viscous dissipation and acoustic
  radiation. Acoustic radiation losses are represented by terms
  involving the (finite) speed of sound $c$ in the
  Keller-Miksis-Parlitz equation. Bubble oscillations can also decay
  due to thermal energy losses, but such damping is neglected in
  comparison with viscous effects \cite{Leighton1994}. While the speed
  of sound is finite in the model employed here it is assumed to be
  sufficiently large so that the phase variation in the incoming
  ultrasound field over the distances of the order of the bubble
  radius are neglected.

In many bubble-acoustic studies, equations are left in dimensional form.
However, to reduce the total number of the governing parameters, we
make equation (\ref{KMPE}) non-dimensional using the equilibrium
radius $R_0$ and the inverse ultrasound frequency $\omega^{-1}$ as the
length and time scales, respectively, so that the equation is
rewritten in terms of non-dimensional bubble radius $r=R(t)/R_0$ and
time $\tau=\omega t$ as
\begin{eqnarray}
 \ddot r[ (1-\Omega \dot r)r+\Omega \R+\S r^2]&=&(\Omega\dot
  r-3)\frac{\dot r^2}{2}-\frac{\W+\R\dot r}{r}\nonumber\\
  &&+(\M+\W)\frac{[1+(1-3\kappa)\Omega\dot
    r]}{r^{3\kappa}}-2\S r\dot r^2\label{KMPEnd}\\ 
  &&-(1+\Omega\dot r)(\M+\Me\sin\tau)-\Me \Omega r\cos
  \tau\,,\nonumber
\end{eqnarray}
where
\begin{eqnarray}
  &\Omega=\frac{\omega R_0}{c}\,,\
  \R=\frac{4\mu}{\rho\omega R_0^2}\,,\
  \W=\frac{2\sigma}{\rho\omega^2R_0^3}\,,&\nonumber\\
  &\M=\frac{P_0-P_v}{\rho\omega^2R_0^2}\,,\
  \Me=\frac{\alpha}{\rho\omega^2R_0^2}\,,\
  \S=\frac{R_0}{s}\,.&\label{ndpar}
\end{eqnarray}
Each of the non-dimensional groups listed above has a straightforward
physical meaning. Since in practical applications the driving
ultrasound frequency is usually fixed, parameter $\Omega$, which is the
ratio of the equilibrium bubble radius and the acoustic wavelength,
characterises the bubble size. Parameters $\R$ and $\W$
characterise  the viscous dissipation and surface tension effects,
respectively. They can be treated as inverse Reynolds
and Weber numbers. 
%{\cmrd In addition to the action of fluid viscosity,
 % bubble oscillations decay due to thermal energy and acoustic radiation losses.   
%Thermal damping is neglected in comparison with viscous damping
%\cite{Leighton1994} while acoustic radiation losses are represented by  
%terms involving the (finite) speed of sound $c$ in the  
%Keller-Miksis-Parlitz equation.}   
Parameter $\M$ represents elastic properties of the
gas and its compressibility, while $\Me$ is the measure of the
external acoustic excitation. Finally, parameter $\S$ is effectively
the inverse distance between the bubble centre and the wall.

\section{Range of parametric values of interest\label{SecPR}}
Fluids where microbubble acoustics is of primary practical
interest are typically water-based. Therefore in order to estimate the
values of the governing non-dimensional parameters we use the
following fluid properties corresponding to water at $20^\circ$C:
$c=1484$  m/s, $\mu=10^{-3}$  kg m/s, $\sigma=7.25\times 10^{-2}$
N/m, $\rho=10^3$  kg/m$^3$, $p_v=2330$ Pa. We also assume that the
gas trapped inside the bubble is air at atmospheric pressure
$P_0=10^5$ Pa and use the value of $\kappa=\frac{4}{3}$ for the
polytropic exponent.

A typical practical range of the microbubble radii is $R_0=0.5-20\
\mu$m. The frequency used in medical ultrasound imaging is 
around $f_e=1$ MHz, and the driving pressure amplitude is in the range
$\alpha=10^5-10^6$ Pa. In this study we consider bubbles that are
assumed to preserve their spherical shape. This introduces the natural
limitation on the value of $s>2R_0$.

\begin{figure}
  \centerline{\def\svgwidth{1.1\textwidth}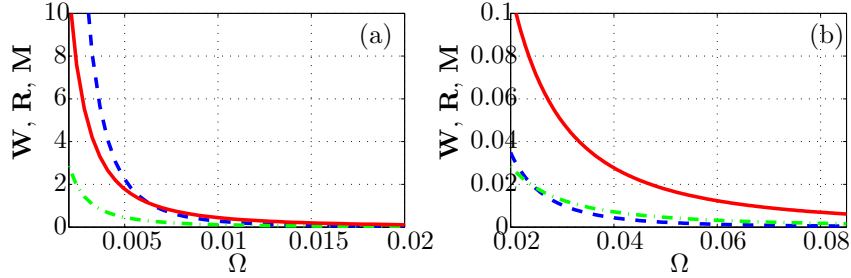}
  \caption{Variation of non-dimensional parameters $\W$ (dash-dotted
    line), $\R$ (dashed line) and $\M$ (solid line)  with the
    non-dimensional bubble radius $\Omega$: (a)~$\Omega<0.02$, (b)
    $\Omega>0.02$.\label{physpar}} 
\end{figure} 
For these physical parameters the values of non-dimensional
groups $\M$, $\W$, $\R$ and $\Omega$ estimated for bubbles of various
radii are linked as
\begin{equation}
  \R=\frac{\R_0}{\Omega^2}\,,\quad
  \W=\frac{\W_0}{\Omega^3}\,,\quad
  \M=\frac{\M_0}{\Omega^2}\,,\quad
  \Me=\frac{\Me_0}{\Omega^2}\,,\label{ndparom}
\end{equation}
where
\begin{eqnarray}
  &\ds\R_0=\frac{4\mu\omega}{\rho c^2}\approx1.14\times10^{-5}\,,\quad
  \W_0=\frac{2\sigma\omega}{\rho c^3}\approx2.79\times10^{-7}\,,\nonumber\\
  &\M_0=\frac{P_0-P_v}{\rho c^2}\approx4.43\times10^{-5}\,,&\label{ndparom0}\\
  &\ds4.5\times10^{-5}\lesssim\Me_0=\frac{\alpha}{\rho
    c^2}\lesssim4.5\times10^{-4},\nonumber&
\end{eqnarray}
for $0.002\lesssim\Omega\lesssim0.085$ (see Fig.~\ref{physpar}),
and $10^5\,{\rm Pa}<\alpha<10^6\,{\rm Pa}$. For
$\Omega\gtrsim0.008$ corresponding to $R_0\gtrsim2\,\mu$m the
magnitude of $\M$ exceeds those of $\W$ and $\R$, see
Fig.~\ref{physpar}(b). This represents the well-known fact that the
dynamics of larger microbubbles is mostly determined by the gas
elasticity, while both viscous dissipation and surface tension
play secondary roles. However for smaller bubbles the value of $\W$
increases rapidly (cubically, see (\ref{ndparom})) and thus bubble
dynamics is influenced by the surface tension to a greater degree. The
role of viscosity for microbubbles of all sizes remains relatively
small.

The other two independent non-dimensional parameters are $\S$ and
$\Me$. The value of $\S$ varies from 0 (isolated bubble far away from
the wall) to 1/2 (bubble near the wall), while typical values of $\Me$
range from 0 to around 0.4 for bubbles greater than about $5\,\mu$m,
but can be of order 10 for bubbles smaller than $2\,\mu$m in diameter.

\section{Small amplitude oscillations\label{secsmlosc}}
When the non-dimensional driving ultrasound pressure amplitude $\Me$
is small, the bubble oscillates near its equilibrium state so that its
instantaneous non-dimensional radius is $r(t)=1+r_1(t)$, where
$r_1(t)\ll1$. Linearizing equation (\ref{KMPEnd}) about $r=1$ we obtain
\begin{eqnarray}
  &(1+\Omega\R+\S)\ddot r_1+(\R+\Omega\A)\dot r_1+\A r_1&\nonumber\\
  &=-\Me(\Omega\cos\tau+\sin\tau)=-\Me\sqrt{1+\Omega^2}\sin(\tau+\phi)\,,& 
\end{eqnarray}
where $\A\equiv(3\kappa-1) \W+3\kappa\M$ and
$\phi=\tan^{-1}\Omega$. The solution of this linear 
equation is
\begin{eqnarray}
  r_1(t)&=&e^{-\frac{\tau}{\tau_0}}(a\sin\omega_0\tau+b\cos\omega_0\tau)
  +a_0\sin(\tau+\phi_0)\label{smas}
\end{eqnarray}
if  $D=(\R-\Omega\A)^2-4(1+S)\A<0$ (this condition is always satisfied
if the viscosity of the fluid remains small as is the case in the
 problem considered). It suggests that the characteristic relaxation
time $\tau_0$
%, $\tau_1$ and $\tau_2$
over which the magnitude of the transient solution reduces by the
factor of $e$, the natural frequency $\omega_0$ of bubble oscillations,
the amplitude $a_0$ and the phase shift $\phi_0-\phi$ of the forced
small amplitude bubble oscillations are given by
\begin{eqnarray}
  \tau_0&=&2\frac{1+\Omega\R+\S}{\R+\Omega\A}\approx
  2\frac{1+\S}{\R_0}\Omega^2\,,\label{tau0}\\ 
 \omega_0&=&\frac{\sqrt{-D}}{2(1+\Omega\R+\S)}\approx
  \sqrt{\frac{\A}{1+\S}}\,,\label{omega0}\\
  a_0&=&\Me\sqrt{\frac{1+\Omega^2}{(1+\Omega\R+\S-\A)^2+(\R+\Omega\A)^2}}
  \approx\frac{\Me}{|1+\S-\A|}\,,\label{a0}\\
  \phi_0-\phi&=&\tan^{-1}\frac{\R+\Omega\A}{1+\Omega\R+\S-\A}
  \approx\tan^{-1}\frac{\R_0}{(1+\S-A)\Omega^2}\,,\label{phi0}
\end{eqnarray}
where
\begin{equation}
  \A\approx
  \begin{cases}
    \ds(3\kappa-1)\frac{W_0}{\Omega^3}\,,&\Omega\ll\Omega_0\,,\\
    \ds3\kappa\frac{M_0}{\Omega^2}\,,&\Omega\gg\Omega_0\,,
  \end{cases}
  \quad\Omega_0=\frac{3\kappa-1}{3\kappa}\frac{\W_0}{\M_0}\approx0.0047\,.\label{Om0}
\end{equation}
The approximate values above are obtained by retaining only the largest
parameters (see Fig.~\ref{physpar} and equations~(\ref{ndparom}) and
(\ref{ndparom0})) entering the expressions. These formulae provide a
straightforward means of identifying the dominant physical processes
taking place in acoustically forced  microbubble oscillations. The
value of $\Omega_0$ determines the bubble size at which the major
physical property determining the characteristics of bubble
oscillations switches from surface tension of the gas/liquid interface
to elasticity of the entrapped gas. As noted earlier, and now quantified 
by (\ref{Om0}), for the considered fluid properties, only very small
bubbles with radii smaller than about $1\,\mu$m would be affected by
the surface tension. We also conclude that the bubble oscillation energy
dissipation rate increases and, subsequently, the relaxation time
decreases due to the action of two physical mechanisms: a viscous
dissipation characterised by $\R$ and losses due to the acoustic
radiation of the bubble, which are proportional to $\kappa\Omega\M$,
yet viscous dissipation always dominates the relaxation process. The
proximity of the wall (e.g.~the increasing value of $\S$)  leads to
an increase in the relaxation time i.e.~to the preservation of the
oscillation energy due to the reflection of acoustic waves from the
wall back towards the bubble. This decrease in damping owing to the 
presence of the wall under linear theory has been noted before 
\cite{Illesinghe2009,Manasseh2009}.
\begin{figure}
  \centerline{\def\svgwidth{\textwidth}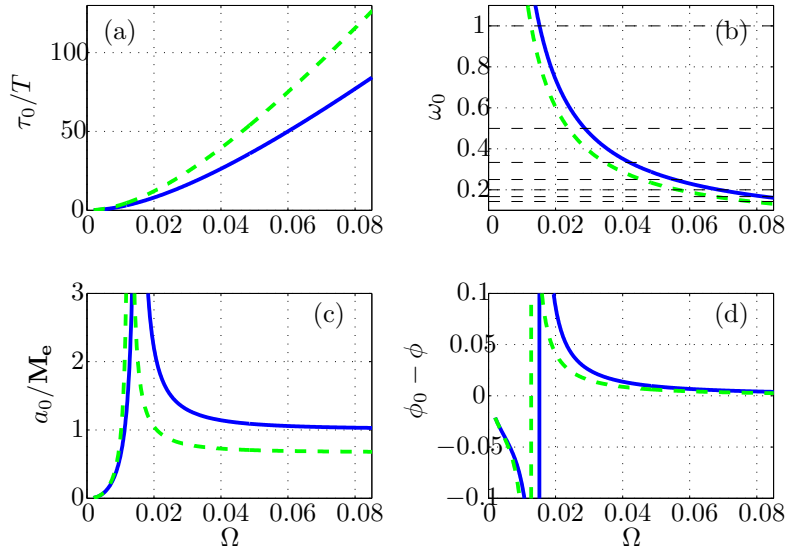}
  \caption{Characteristics of small amplitude oscillations as
    functions of the non-dimensional bubble radius
    $\Omega$ for a bubble away ($\S=0$, solid lines) and near the wall
    ($\S=\frac{1}{2}$, dashed lines). Intersections of the horizontal
    dashed lines with the $\omega_0$ lines in plot (b)
    correspond to subharmonic resonance values of
    $\Omega$ listed in Table~\ref{restbl}. \label{smlosc}} 
\end{figure}

As mentioned in Section \ref{SecPR}, the natural frequency of
small-amplitude bubble oscillations at $\Omega>\Omega_0$ (this
condition is satisfied for almost a complete range of bubble sizes
used in practice) is predominantly determined by 
the gas elasticity with the surface tension playing a negligible
role. The amplitude of oscillations is directly proportional to
that of the ultrasound forcing $\Me$ and decreases due to the
restricting influence of the wall as $\S$ increases (when the bubble
approaches the wall). The phase lag $\phi_0-\phi$ between the forcing
and the bubble response is mostly due to the effects of fluid
viscosity  and acoustic radiation. The proximity of the wall forces a
bubble to follow the external excitation more closely.

The variations of characteristics of small-amplitude oscillations with
the non-dimensional bubble radius are presented in
Fig.~\ref{smlosc}. The overall trends are that the relaxation time
increases and the natural frequency of bubble oscillations decreases
with the bubble radius. From a computational point of view it is
important to note that depending on the initial conditions and the
bubble equilibrium size it could be required to integrate the
governing equations over more than 300 forcing periods before a
statistically steady solution can be established for a bubble located
away from the wall. For a near-wall bubble the transient time can be
greater by at least a factor of two.

One of the most important sources of information regarding the
anticipated bubble behaviour is the bubble natural frequency diagram
shown in Fig.~\ref{smlosc}(b). The value of $\omega_0$ monotonically
decreases with the non-dimensional bubble radius $\Omega$ and becomes unity at
\begin{equation}
  \Omega=\Omega_1\approx\sqrt{\frac{3\kappa\M_0}{1+\S}}
\end{equation}
(see Table~\ref{restbl} for numerical values) that corresponds to the
size of the bubble at which it resonates strongly with the external
forcing. Typical resonance features such as a rapid increase of the
bubble oscillation amplitude up to the maximum value 
\begin{equation}
  a_{0\max}\approx\frac{\Omega_1^2\Me}{\R_0}
  =\frac{3\kappa\M_0\Me}{(1+\S)\R_0}
\end{equation}
and the switch of the phase lag
$\phi_0-\phi$ from $\frac{\pi}{2}$ to $-\frac{\pi}{2}$ are observed at 
this point, see Fig.~\ref{smlosc}(c, d). For $\Omega<\Omega_1$ the
amplitude of bubble oscillations decreases very rapidly (see
Fig.~\ref{smlosc}(c)). A bubble effectively stops responding to an
incoming ultrasound wave and becomes ``invisible'' in acoustic imaging
applications. Therefore in our further analysis we focus on the
nontrivial bubble behaviour observed at $\Omega>\Omega_1$. We also
note that the small amplitude oscillations of a bubble near the wall
are similar to those of a distant bubble, however the resonance shifts
towards smaller values of $\Omega_1$  (see Table~\ref{restbl}). This
frequency shift has been also noted in \cite{Payne2005} and the
potential value of a frequency shift in identifying targeted
microbubbles via a filtering approach has been recognised in
\cite{Payne2011}. Therefore it is expected that the acoustic signature
of initially ``silent'' small bubbles will become more pronounced as
they approach the wall. This feature can potentially be used for
estimating the likelihood of bubbles reaching the walls of blood
vessels in applications such as targeted drug delivery and ultrasound imaging.

The fact that the non-dimensional natural frequency of bubble
oscillations remains smaller than 1 for a large portion of the practical bubble size
range suggests that non-linear subharmonic resonances can occur. The
intersections of the horizontal dashed lines with the $\omega_0$
curves in Fig.~\ref{smlosc}(b) define the subharmonic resonant
values of $\Omega$ which are listed in Table~\ref{restbl}. Of course, 
the subharmonic resonances can only exist owing to nonlinear effects manifested
when the amplitude is no longer small. As will be
shown in the subsequent sections, the presence of subharmonic
resonances defines what type of oscillations are observed.
Finally, we note that larger distant and near-wall bubbles follow the
external forcing quite closely so that the phase lag remains close to
zero away from the main resonance.
\begin{table}
\caption{The values of parameter $\Omega$ at which subharmonic
  resonances with frequencies listed in the top row are expected to
  occur for a bubble away (middle row) and near the wall (bottom
  row).\label{restbl}}
\renewcommand{\arraystretch}{1.75}
\centerline{
  \begin{tabular}{|c|c|c|c|c|c|c|c|}
    \hline
    $\omega_0$&1&$\frac{1}{2}$&$\frac{1}{3}$&$\frac{1}{4}$
    &$\frac{1}{5}$&$\frac{1}{6}$&$\frac{1}{7}$\\
    \hline
    $\Omega$ ($S=0$)&0.0152&0.0287&0.0421&0.0554&0.0688&0.0821&0.0955\\
    \hline
    $\Omega$ ($S=\frac{1}{2}$)&0.0128&0.0238&0.0348&0.0456&0.0566&0.0674&0.0784\\
    \hline
    \end{tabular}}
\end{table}

\section{Finite amplitude periodic solutions}
\subsection{Floquet analysis}
When the forcing amplitude $\Me$ increases, the bubble oscillation
amplitude becomes large and solutions discussed in the
previous section are replaced with non-linear solutions
$r_0(\tau)$. 
%{\cmrd This is irrelevant: Floquet theory has nothing to do with the
%wall and it has not be used before in theses references. As noted
%earlier, non-linear dynamical-systems analyses  
%have been computed before for isolated acoustically-driven microbubbles 
%\cite{Lauterborn1976,Parlitz1990}, but now we are able to consider the
%wall effect.} 
The solutions are not sinusoidal anymore, yet are still
$T$-periodic, where $T=2\pi$ is the period of the external forcing.
However this remains true only up to a certain critical value
of the forcing amplitude at which the $T$-periodic solution
undergoes a bifurcation to a different state. To determine the exact
parametric values for  the bifurcation point and the nature of the
bifurcation we look for a solution in the form $r(t)=r_0(\tau)+r'(\tau)$,
where we assume that $r'(\tau)\ll r_0(\tau)$ over the period of one forced
oscillation. Numerically, $r_0(\tau)$ is obtained by solving equation
(\ref{KMPEnd}) with the periodic boundary conditions $r_0(0)=r_0(T)$
and $\dot r_0(0)=\dot r_0(T)$ (see Appendix~\ref{appendix} for details
of numerical implementation). To investigate the stability of such a
solution we consider the linearization of equation (\ref{KMPEnd})
about $r_0(\tau)$
\begin{alignat}{2}
    [ (1-\Omega &\dot r_0)r_0+\Omega \R+\S r_0^2]\ddot r'=&&\nonumber\\
    &-\Bigg[3\dot r_0+\frac{\R}{r_0}+4\S r_0\dot r_0
    +\Omega\Bigg(\M-\frac{3}{2}\dot r_0^2-R_0\ddot r_0 
      +\frac{(3\kappa-1)(\M+\W)}{r_0^{3\kappa}}
      +\Me\sin t\Bigg)\Bigg]\dot r'&&\label{eqfloq}\\
    &+\Bigg[\frac{\W+r_0\dot r_0}{r_0^2}-(1-\Omega\dot r_0)\ddot
    r_0-\frac{3\kappa(\M+\W)(1+(1-3\kappa)\Omega\dot
      r_0)}{r_0^{3\kappa+1}}-\Me\Omega\cos t-2\S(\dot r_0^2+r_0\ddot
    r_0)\Bigg]r'\,.\nonumber&&
\end{alignat}
and apply Floquet analysis \cite{Nayfeh1995}. In brief, we
solve equation (\ref{eqfloq}) with known periodic coefficients
depending on $r_0(\tau)$ and $\dot r_0(\tau)$ over the interval
$\tau\in[0,T]$ subject to two sets of linearly independent initial
conditions $[r'_1(0),\dot r'_1(0)]=[1,0]$ and $[r'_2(0),\dot
r'_2(0)]=[0,1]$. The values of the obtained solutions at $\tau=T$ form
a monodromy matrix $Y$
\begin{equation}
  Y=\left[\begin{array}{cc}
      r'_1(T)&r'_2(T)\\
      \dot r'_1(T)&\dot r'_2(T)
    \end{array}\right],
\end{equation}
whose (generally complex) eigenvalues
$\sigma_{1,2}=\sigma_{1,2}^R+i\sigma_{1,2}^I=|\sigma_{1,2}|e^{i\theta_{1,2}}$
are Floquet multipliers. According to the Floquet theorem, the solution of
equation (\ref{eqfloq}) satisfies the following relationship
\begin{equation}
  [r'(nT),\dot r'(nT)]=\sigma^n[r'(0),\dot
    r'(0)]=|\sigma|^ne^{in\theta}[r'(0),\dot r'(0)]\,.
\end{equation}
Therefore the periodic solution $ r_0(\tau)$ is unstable if the
magnitude of at least one of the Floquet multipliers, $\max
|\sigma_{1,2}|$, exceeds unity. At the bifurcation point we must have
$\max|\sigma_{1,2}|=1$. This condition determines neutral stability of
a periodic solution with respect to infinitesimal disturbances and the
bifurcation type is determined by the complex value of the Floquet
multiplier with the unit magnitude. For example, if
$\sigma=e^{2i\pi/n}$ so that $\sigma^n=\sigma^{2i\pi}=1$ then the
solution of (\ref{eqfloq}) will be repeated for the first time after
$n$ forcing periods $T$:
\[r'(nT)=r'(0)\,,\quad {\dot r}'(nT)={\dot r}'(0)\,,\quad
r'(mT)\ne r'(0)\,,\quad {\dot r}'(mT)\ne{\dot r}'(0)\,,\]
where $m<n$, meaning that $T\to nT$ bifurcation has occurred. In
particular, in the situation when one of the Floquet multipliers
becomes equal to $e^{i\pi}=-1$, i.e.~when $n=2$, a period-doubling
bifurcation is observed.

\subsection{Medium amplitude oscillations}
\begin{figure}
 \centerline{ \def\svgwidth{1.1\textwidth}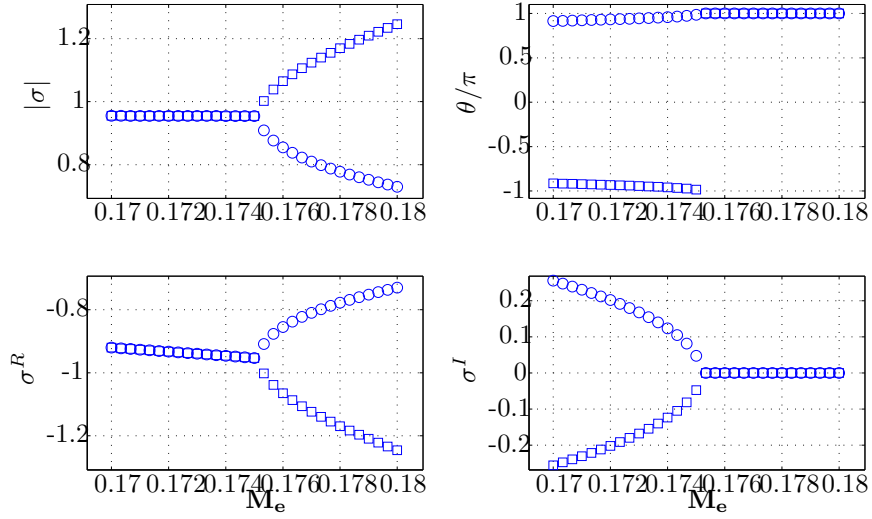}
  \caption{Floquet multipliers for medium amplitude bubble oscillations
    away from the wall ($\S=0$) at $\Omega=0.0423$. Circles and
    squares correspond to two distinct multipliers. Circles and
    squares correspond to two distinct multipliers.\label{egvR10_5S0T2}}
\end{figure}
The typical behaviour of Floquet multipliers as the forcing amplitude
$\Me$ increases is illustrated in Fig.~\ref{egvR10_5S0T2}. The
magnitude of one of the multipliers increases above the value of $1$
while the second multiplier remains smaller than unity. The phase of
the larger of the two Floquet multipliers becomes $\theta=\pi$ meaning
that $n=\frac{2\pi}{\theta}=2$ and thus the period-doubling
bifurcation occurs at $\Me=\Me_{\,cr}$. Note that a bifurcation of
Floquet multipliers seen in Fig.~\ref{egvR10_5S0T2} is not related to
the period-doubling bifurcation of the solution of (\ref{KMPEnd}); the
latter is determined only by the fact that one of the multipliers becomes
equal to $-1$ regardless of whether $\sigma$ itself undergoes any
bifurcations. Thus the parametric location of a bifurcation point is
accurately determined by Floquet multipliers that change continuously
with $\Me$ with one of them becoming equal to $-1$ at the
critical value of $\Me=\Me_{\,cr}$.

The period-doubling bifurcations are detected in small to medium
amplitude oscillations for all considered values of non-dimensional
bubble radius $\Omega$. Numerical experiments confirm that the
transition between $T$- and $2T$-periodic solutions occurs at the same
value of $\Me=\Me_{\,cr}$ regardless of whether $\Me$ is gradually
increased or decreased. Since no hysteresis is observed we conclude
that microbubbles undergo supercritical period-doubling bifurcation in
all considered regimes.
\begin{figure}
  \centerline{\def\svgwidth{\textwidth}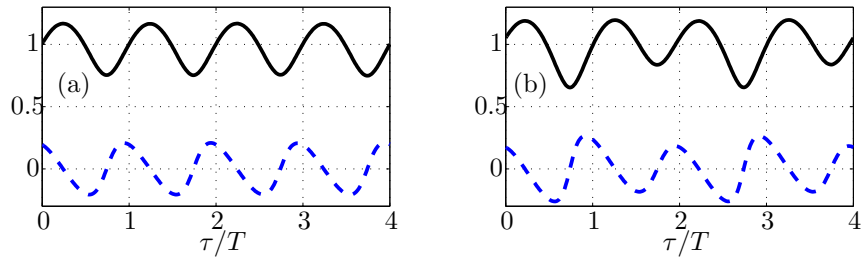}
  \caption{Bubble oscillations away from the wall ($\S=0$) at
    $\Omega=0.0423$ (a) before ($\Me=0.17$) and (b) after ($\Me=0.18$)
    the period-doubling bifurcation. The solid and dashed lines
    represent $r(\tau)$ and $\dot r(\tau)$,
    respectively.\label{M1718S00}}
\end{figure}
\begin{figure}
  \centerline{\def\svgwidth{\textwidth}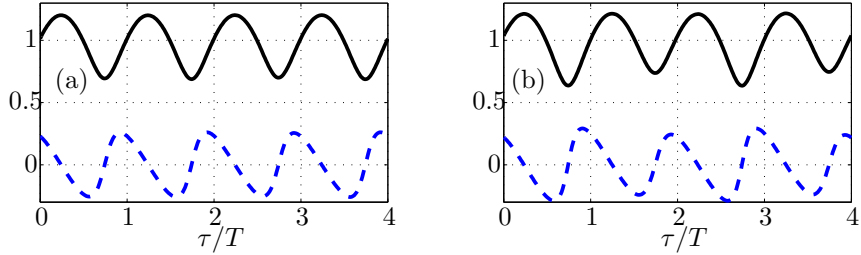}
  \caption{Oscillations of a bubble near the wall
    $\left(\S=\frac{1}{2}\right)$ at $\Omega=0.0423$
    (a) before ($\Me=0.32$) and (b) after ($\Me=0.33$) 
    the period-doubling bifurcation. The solid and dashed lines
    represent $r(\tau)$ and $\dot r(\tau)$,
    respectively.\label{M3233S05}} 
\end{figure}

The examples of bubble oscillations for $\Omega=0.0423$ corresponding
to the equilibrium bubble radius $R_0=10\,\mu$m before and after the
period-doubling bifurcations occurring at $\Me_{\,cr}\approx0.17532$
for a bubble away from the wall and at $\Me_{\,cr}\approx0.32621$ for
a bubble near the wall are shown in Figs~\ref{M1718S00} and
\ref{M3233S05}. While the qualitative behaviour of the distant and
near-wall bubbles remains the same we note that the proximity of
the wall causes a significant increase in the ultrasound pressure
required to induce a supercritical period-doubling
bifurcation. Equivalently, this means that the presence of the wall
has a strong stabilising influence on the oscillations of bubbles with
the equilibrium radius $R_0\gtrsim10\,\mu$m. The pressure amplitude that 
would have to be applied to cause period doubling is nearly doubled by 
proximity to the wall. In clinical terminology, this implies that an ultrasound 
scanner would have to be set to produce twice the ``mechanical index''
in order to see this non-linear response at the wall, a very significant
change. However, we will show in the following sections that for
smaller bubbles this trend is reversed.

\subsection{Large amplitude oscillations\label{seclao}}
\begin{figure}
  \centerline{\def\svgwidth{1.1\textwidth}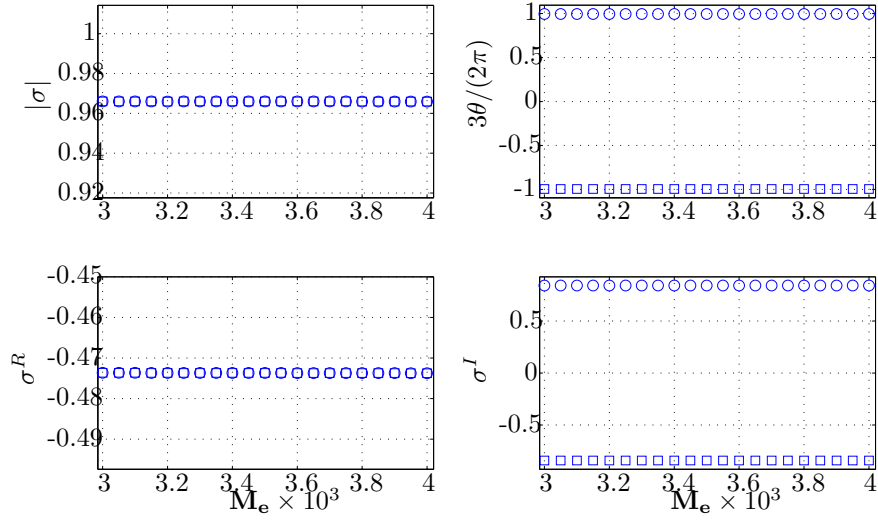}
  \caption{Floquet multipliers for stable $T$-periodic bubble
    oscillations away from the wall ($\S=0$) at
    $\Omega=0.0423$ indicating the presence of $3T$-periodic
    oscillations. Circles and squares correspond to two distinct
    multipliers.\label{egvR10_5S0T3}}
\end{figure}
\begin{figure}
  \centerline{\def\svgwidth{.9\textwidth}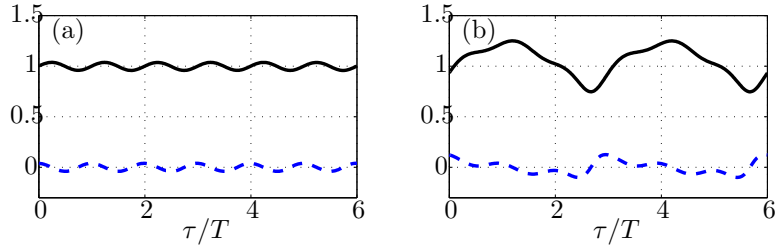}
  \caption{(a) Small and (b) large amplitude oscillations of a bubble
    away from the wall ($\S=0$) at $\Omega=0.0423$ and
    $\Me=0.035$. The solid and dashed lines represent $r(\tau)$ and
    $\dot r(\tau)$, respectively. \label{M035S00}}
\end{figure}
Away from the parametric location of the period-doubling bifurcation 
the magnitudes of Floquet multipliers computed for small amplitude
$T$-periodic solutions (that are accurately approximated by
(\ref{smas})) for a bubble located far away from the wall remain smaller
than unity, see Fig.~\ref{egvR10_5S0T3}. This means that such
oscillations are stable and the bubble can follow external ultrasound
excitation for an indefinitely long time (see
Fig.~\ref{M035S00}(a)). However, the phase $\theta$ of the Floquet
multipliers provides very remarkable information. It turns out that
$\theta_{1,2}$ asymptotes to the values $\pm\frac{2\pi}{3}$. Thus
$n=\frac{2\pi}{\theta_{1,2}}=\pm3$. Even though no bifurcation is
detected, since $|\sigma_{1,2}|<1$, the Floquet multipliers strongly
indicate that a $3T$-periodic solution may exist. However, in contrast
to the $2T$-periodic solution, it is disjoint from the $T$-periodic
solution in that sense that it cannot be obtained from it via a
continuous variation of physical governing parameters such as $\Omega$
or $\Me$ alone.

A careful numerical investigation reveals that the governing equation
(\ref{KMPEnd}) indeed admits large amplitude oscillation solutions
that are not described by (\ref{smas}) even when the external forcing
is weak. Consistent with the predictions based on Floquet analysis,
such large amplitude oscillations have a larger period of $3T$ as is
confirmed by Fig.~\ref{M035S00}(b). In particular, for the bubble of
equilibrium radius $R_0=10\,\mu$m ($\Omega=0.0423$), the minimum value
of the forcing amplitude at which $3T$-periodic solution is still detected
numerically is found to be $\Me\approx0.0342$ (corresponding to
$\alpha\approx135$ kPa) which agrees well with the value estimated
from Fig.~10(a) in \cite{Chong2010}.

\begin{figure}
  \centerline{\def\svgwidth{1.2\textwidth}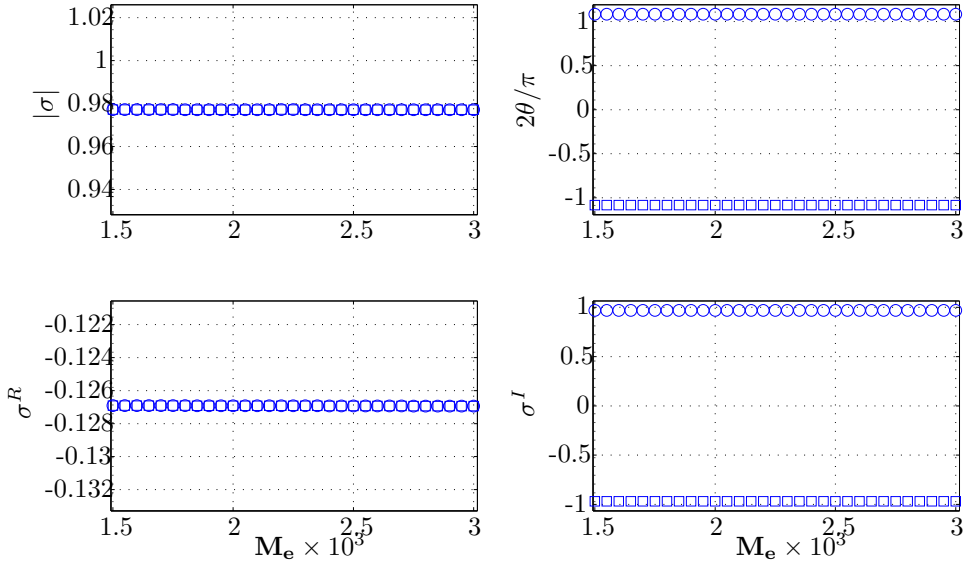}
  \caption{Floquet multipliers for stable $T$-periodic bubble
    oscillations of the bubble near the wall ($\S=\frac{1}{2}$)
    at $\Omega=0.0423$ indicating the presence of $4T$-periodic
    oscillations. Circles and squares correspond to two distinct
    multipliers.\label{egvR10_5S5T4}}
\end{figure}

\begin{figure}
  \centerline{\def\svgwidth{\textwidth}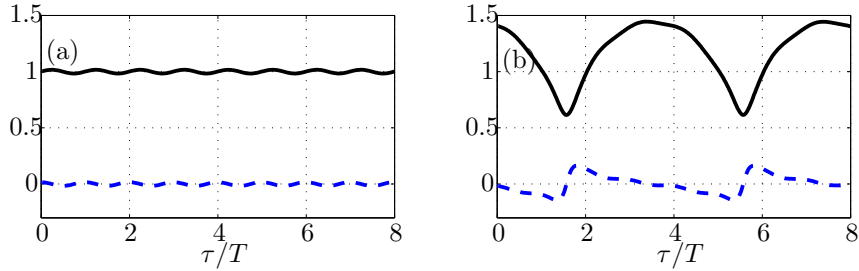}
  \caption{(a) Small and (b) large amplitude oscillations of a
    bubble near the wall $\left(\S=\frac{1}{2}\right)$ at
    $\Omega=0.0423$  and  $\Me=0.023$. The solid and dashed lines
    denote $r(\tau)$ and $\dot r(\tau)$, respectively. \label{M023S05}}
\end{figure}

Similarly, Floquet multipliers computed away from a period-doubling
bifurcation for a bubble near the wall (see
Fig.~\ref{egvR10_5S5T4}), confirm that small amplitude $T$-periodic
solutions are stable, see Fig.~\ref{M023S05}(a). However
in contrast to the distant bubble case, the phase $\theta$ of Floquet
multipliers asymptotes to the value of $\theta_{1,2}=\pm\frac{\pi}{2}$
so that $n=\frac{2\pi}{\theta}=4$ indicating the presence of
$4T$-periodic solutions disjoint from the $T$-periodic
oscillations. Such solutions were indeed found numerically, see
Fig.~\ref{M023S05}(b).
Thus we conclude that the period of large amplitude bubble
oscillations depends on the proximity of the bubble to the wall and
increases from $3T$ far away from the wall to $4T$ next to it.
Floquet stability analysis (not detailed here) was performed for the
$3T$- and $4T$-periodic large amplitude oscillations; in addition,
equation (\ref{KMPEnd}) was solved numerically over several hundred
forcing periods. This confirmed that the $3T$- and $4T$-periodic
oscillations are stable and thus can co-exist with small amplitude
$T$-periodic oscillations.

Having established that the small and large amplitude oscillations
illustrated in Figs~\ref{M035S00} and \ref{M023S05} cannot be obtained
from each other by a parametric continuation, we ask a natural question:
what defines the type of the observed oscillations in practice?
Numerical experiments show that it depends not on the magnitude of the
forcing, but rather on the initial conditions: large amplitude
oscillations are typically established if a sufficiently large value
of $\dot r(0)$ is specified. Physically, this corresponds to a
pressure impulse that causes microbubble to initially contract with a
large speed and then relax to large amplitude long-period
oscillations. This is of particular practical significance since 
clinical ultrasound scanners do not apply continuous forcing, but employ a 
series of discrete pulses.

These numerical results emphasise that initial excitation can have a
long-term (in fact, permanent and determining) effect on the observed
bubble oscillation patterns. Therefore the results of the analysis of the
bubble dynamics based on numerically obtained bifurcation diagrams
(Poincar\'e maps) that  have been a popular tool of bubble
dynamics analysis \cite[e.g.]{Parlitz1990,Macdonald2006,Chong2010} have to be
interpreted  carefully. In obtaining such diagrams it is sometimes
assumed that the initial conditions are ``fully forgotten'' once the
transients have decayed and a periodic solution has been
established. However this might not be the case due to the existence
of multiple solutions of a highly non-linear system (\ref{KMPEnd})
each having its own basin of attraction \cite{Parlitz1990}. Thus the
initial conditions should be added to the set of parameters
characterising the solutions along with those given by
(\ref{ndpar}). Failure to do so may lead to an ambiguous
interpretation of the computed bubble dynamics.

\section{Bubble oscillation maps}

\subsection{Bubble far from walls}

\begin{figure}
  \centerline{\def\svgwidth{.8\textwidth}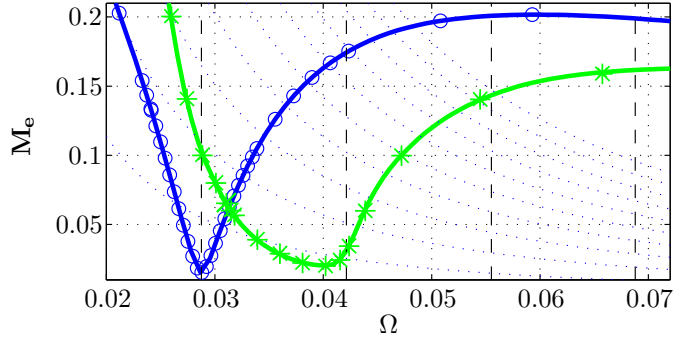}
  \caption{Oscillation map for a bubble away from the wall ($\S=0$). The
    parametric boundaries for period-doubling and tripling transitions
    are shown by circles and stars, respectively. Vertical dashed lines
    show the estimated positions of subharmonic resonances listed in the
    middle row of Table~\ref{restbl}. The dotted curves show
    isocontours $\alpha=$const.~with the values of $\alpha$ ranging
    (from left to right) from 100 to 1000 kPa at 100 kPa
    increments.\label{bifdiagS0}}
\end{figure}
Various types of oscillations that can be experienced by a microbubble
located far away from a wall are summarised in
Fig.~\ref{bifdiagS0}. The line marked by circles is the parametric
locus of a period-doubling bifurcation: $T$-periodic oscillations are
stable below this line and stable $2T$-periodic solutions replace them
above it. The values of $\Me$ corresponding to the period-doubling
bifurcation valid at least up to 4 significant digits were obtained
iteratively by systematically applying the Floquet analysis implemented as
described in the Appendix. 

The $2T$-periodic solution boundary has the characteristic
minimum at $(\Omega,\Me)\approx(0.029,0.015)$ below which no bubbles
can oscillate with the half-frequency of the driving ultrasound
wave. Note that the above value of $\Omega$ virtually coincides with
that at which the natural frequency $\omega_0$ of small amplitude
oscillations becomes equal to $\frac{1}{2}$ (see Fig.~\ref{smlosc} and
Table~\ref{restbl}). Therefore the minimum of the driving pressure
amplitude resulting in the transition to $2T$ periodic solution is
related to the first subharmonic resonance observed in the system,
an observation consistent with conclusions regarding the role of
resonances in the bubble dynamics discussed in \cite{Parlitz1990}. The
existence of this minimum could also be qualitatively related to the
amount of energy required to initiate and maintain oscillations for
various values of $\Omega$. It is intuitively clear that the energy
required to start oscillations (which is proportional to the square
of the oscillation amplitude and thus to the square of the
characteristic bubble radius) decreases with the bubble size and so
does $\Me_{\,cr}$. However, as evidenced by Fig.~\ref{physpar} for
small values of $\Omega$, the value of $\M$, the parameter
characterising acoustic losses, increases rapidly, and so does the
bubble energy loss rate (see the decrease in the relaxation time
$\tau_0$ in Fig.~\ref{smlosc}(a) that is inversely proportional to the
energy loss rate). To compensate for this enhancement of the
oscillation energy loss due to acoustic radiation, the influx of the
ultrasound energy has to increase in order to maintain
oscillations. Thus the value of $\Me_{\,cr}$ must increase 
for small bubbles. The minimum in the $\Me_{\,cr}$ curve thus
corresponds to the bubble size that insures the minimum overall
oscillation energy loss due to viscous dissipation and acoustic
radiation.
  
Note that since $\displaystyle \Omega^2\Me=\frac{\alpha}{\rho c^2}$
we conclude that no bubble can produce a half-frequency response if
\[\alpha<\alpha_{2T}\approx1.26\times10^{-5}\rho
c^2\approx27.8\mbox{ kPa}\,.\]
Therefore, setting $\alpha\approx\alpha_{2T}$ in an experiment and
sweeping through a range of frequencies $f_e$, one can detect the
value of frequency $f_{2T}$ at which the half-frequency response is
first heard. This would provide a straightforward estimation of the
size of the bubble away from the wall
\[R_0\sim 0.029\frac{c}{2\pi f_{2T}}\approx\frac{6.9}{f_{2T}}\,,\]
where $f_{2T}$ is in MHz and $R_0$ is in $\mu$m. For example, for
the considered frequency $f_e=1$MHz bubbles for which the
half-frequency response would occur first would have the equilibrium
radius $R_0\approx6.9\,\mu$m.

The line marked by stars in Fig.~\ref{bifdiagS0} represents the
minimum non-dimensional amplitude of the ultrasound forcing at which
stable large amplitude $3T$-periodic oscillations were  still detected
numerically. These solutions, while remaining stable, cease to exist in
a catastrophic way (resembling a fold bifurcation) when the value of
$\Me$ is gradually decreased below the starred line. The smallest
values of $\Me$ for which $3T$-periodic solutions were still detected
were determined up to 3 significant digits using the parametric
continuation procedure detailed in the Appendix.

The resulting parametric boundary for the existence of $3T$-periodic
solutions has a minimum at
$(\Omega,\Me)\approx(0.040,0.020)$. Again the parametric location of 
this minimum is close to the value of $\Omega$ where the bubble natural
frequency $\omega_0$ becomes equal to $\frac{1}{3}$ of the
driving frequency, see Fig.~\ref{smlosc}(b), and
thus it is closely linked to the occurring  subharmonic resonance. A
slight difference between the $\Omega$ location of the minimum and the
resonance value reported in Table~\ref{restbl} is attributed to a
nonlinear shift of the natural oscillation frequency occurring for
finite amplitude oscillations. Based on the value of
$\Me$ at the minimum we conclude that $3T$-periodic oscillations can
only be maintained for
\[\alpha>\alpha_{3T}\approx3.2\times10^{-5}\rho
c^2\approx70.5\mbox{ kPA}\]
and the bubble size estimation formula based on the period tripling
frequency $f_{3T}$ becomes
\[R_0\sim 0.040\frac{c}{2\pi f_{3T}}\approx\frac{9.5}{f_{3T}}\,,\]
where $f_{3T}$ is in MHz and $R_0$ is in $\mu$m. However, given
that $3T$-periodic solutions usually require initial pressure pulse to
induce, using this correlation in practical experiments may be less
convenient than that for $2T$-periodic oscillations.

\begin{figure}
  \centerline{\def\svgwidth{\textwidth}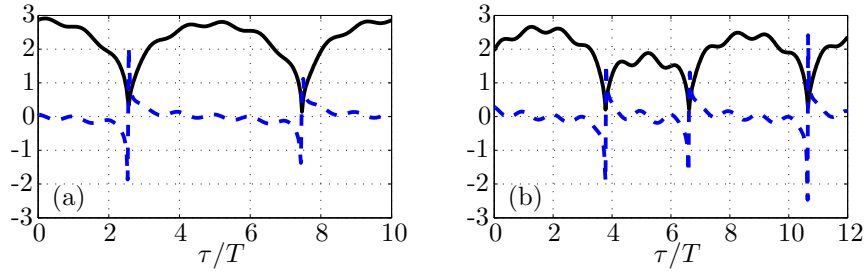}
  \caption{Combined bubble oscillations away from the wall ($\S=0$)
    at $\Omega=0.0423$ for (a) $\Me=0.18$ and (b) $\Me=0.25$. The
    solid and dashed lines represent $r(\tau)$ and $\dot r(\tau)$,
    respectively.\label{M1825S00}}
\end{figure}
The $2T$- and $3T$-periodic solution boundaries intersect at
$(\Omega,\Me)\approx(0.0313,0.0626)$. Therefore it is expected that
as the driving pressure amplitude increases the oscillations of
smaller bubbles to the left of the intersection point will always
experience period-doubling before $3T$-periodic oscillations can be
observed. This fact was also illustrated by Figs~10(a) and 12(a) in
\cite{Chong2010}. The situation is much more complicated for larger
bubbles. For small forcing amplitudes, in the region below the line
marked by stars in  Fig.~\ref{bifdiagS0}, only small amplitude
$T$-periodic oscillations can be observed. In the region between the
lines marked by stars and circles, both stable $T$ and $3T$ periodic
oscillations can exist, and the type of oscillations established
is determined by the initial conditions. Above the curve marked by
circles, $T$-periodic solutions become unstable and cannot be
observed. They are replaced by stable $2T$-periodic oscillations, see
Fig.~\ref{M1718S00}. Again, depending on the initial
conditions the $3T$-periodic solutions also can exist in this region
as was observed in our numerical computations just above the $2T$
transition boundary. However, for larger values of $\Me$, combination
solutions involving both $3T$ and $2T$ components are also
observed. In some instances such solutions appear as $5T$-periodic 
oscillations, see Fig.~\ref{M1825S00}(a). Note that the
physical parameters for which this plot is generated are identical to
those chosen for Fig.~\ref{M1718S00}(b), yet the long-term
solutions obtained for different initial conditions differ drastically.  
In other cases the strongly non-linear interaction between these large
amplitude solutions leads to what appears to be aperiodic
oscillations, see Fig.~\ref{M1825S00}(b), frequently
referred to as chaotic behaviour in literature. Thus we can state that
the parametric region above both curves in  Fig.~\ref{bifdiagS0} is
where chaotic bubble oscillations can potentially be observed. This
region protrudes towards smaller forcing amplitudes near the
intersection of the two curves. Therefore the curve intersection point
defines the size of the bubbles that are most likely to enter the
chaotic regime away from the wall. For $f_e=1$ MHz  it is
$R_0\approx7.4\,\mu$m.

The higher order subharmonic resonances that are expected to exist for
a bubble away from the wall near other vertical dashed lines in
Fig.~\ref{bifdiagS0} did not seem to lead to the appearance of
$nT$-periodic ($n>3$) oscillations at least for the range of
parameters in the figure. We chose not to look for such solutions at
larger values of $\Me$, since they would be far outside the practical
range of ultrasound amplitudes that corresponds to the region between
dotted curves in Fig.~\ref{bifdiagS0}. 

\subsection{Bubble close to a wall}

\begin{figure}
  \centerline{\def\svgwidth{1.2\textwidth}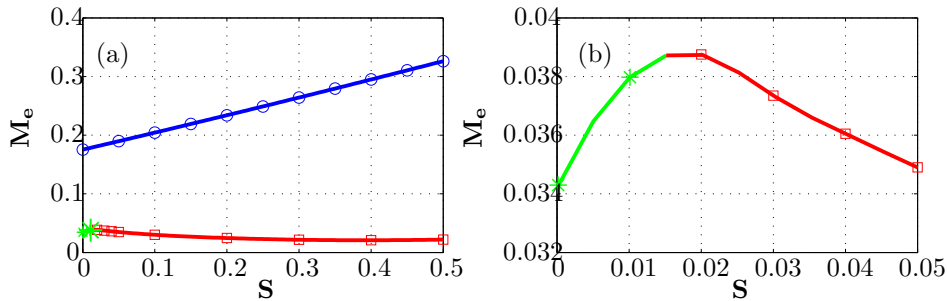}
  \caption{Oscillation map for a bubble approaching the wall for
    $\Omega=0.0423$: (a) complete diagram, (b) close
    up for distant bubbles (computational points are shown by symbols,
    the lines represent a spline interpolation of the data). The
    parametric boundaries for period-doubling, tripling and quadrupling
    transitions are shown by circles, stars and squares,
    respectively.\label{SR10_5}}
\end{figure}
To investigate the effect of the wall proximity on bubble dynamics, we
consider an oscillation map produced for $\Omega=0.0423$
(corresponding to $R_0=10\,\mu$m), with $\S$ varying from 0 (bubble
away from the wall) to $\frac{1}{2}$ (bubble near the
wall). The results are presented in Fig.~\ref{SR10_5}. The only effect
the wall proximity has on the bubble's tendency to undergo a period
doubling bifurcation is quantitative: an essentially linear increase
of the critical value of $\Me$ approximately given by
\[\Me_{\,cr}=\Me_{\,cr0}+d\S\,.\]
For $\Omega=0.0423$ and other parameters corresponding to this
value ($\R=6.4\times10^{-3}$, $\W=3.7\times10^{-3}$,
$\M=2.47\times10^{-2}$, see Fig.~\ref{physpar})
$\Me_{e\,cr0}\approx0.0175$ and $d\approx0.3$.
It was noted in the comparison of Figs~\ref{M1718S00} and
\ref{M3233S05} that
 the presence of the wall can delay the transition to a doubly-periodic 
regime by a factor of almost two in forcing amplitude. This
is intuitively expected as the wall is modelled by introducing an
identical bubble image so that when the bubble approaches the wall the
external ultrasound radiation effectively has to drive two bubbles
instead of one.

In contrast to small amplitude solutions, strongly non-linear large
amplitude oscillations undergo a qualitative change in the vicinity of
the wall: the period of such oscillations changes from $3T$ away from
the wall to $4T$ when the bubble approaches the wall, see lines marked
by stars and squares in Fig.~\ref{SR10_5}. Such a switch occurs  between
$\S=0.01$ and $\S=0.02$, see Fig.~\ref{SR10_5}(b) i.e.~when the bubble
is between 25 and 50 radial distances away from the wall. Therefore the
influence of the wall on the large amplitude oscillations is far-reaching. 
Another important observation is that the forcing amplitude
$\Me$ decreases along the boundary of $4T$-periodic oscillations as
the bubble approaches the wall. Therefore it becomes easier to induce
such oscillations near the wall. Both these facts suggest that the
appearance of the quarter-frequencies in the bubble acoustic response
spectra can be used in practice as an indication of the bubble's
approach to the wall.

\begin{figure}
  \centerline{\def\svgwidth{.8\textwidth}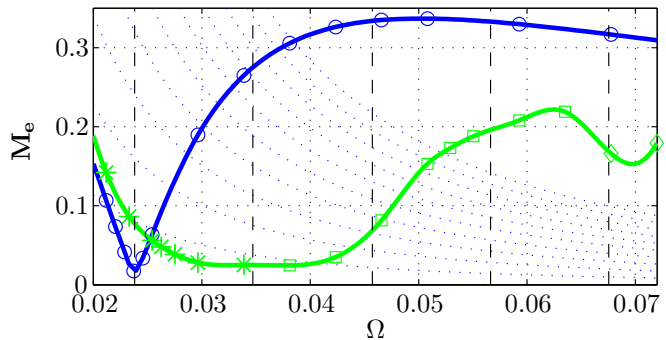}
  \caption{Oscillation map for a bubble near the wall
    ($\S=\frac{1}{2}$). The parametric boundaries for transitions to
    $2T$-, $3T$-, $4T$- and $5T$-periodic oscillations are shown by
    circles, stars, squares and diamonds, respectively. Vertical dashed
    lines show the estimated positions of subharmonic resonances listed
    in the bottom row in Table~\ref{restbl}. The dotted curves show
    isocontours $\alpha=$const.~with the values of $\alpha$ ranging
    (from left to right) from 100 to 1000 kPa at 100 kPa
    increments.\label{bifdiagS5}}
\end{figure}
In Fig.~\ref{bifdiagS5} we present the parametric boundaries
for transitions to various long-period oscillations of a near-wall
bubble. The comparison with Fig.~\ref{bifdiagS0} shows that the
proximity of the wall leads to significantly more complicated bubble
dynamics characterised by the appearance of different oscillation
modes that depend sensitively on the non-dimensional bubble radius
$\Omega$. The shape of the parametric line showing the locations of
period-doubling bifurcations remains similar to that seen in
Fig.~\ref{bifdiagS0} for a bubble away from the wall, with the minimum
detected at the first subharmonic resonance (see the bottom row in
Table~\ref{restbl}). However the figure shows that the behaviour of
bubbles of different sizes as they approach the wall is not the
same. Namely, the excitation amplitude at which the period doubling is
observed for bubbles with a non-dimensional radius
$\Omega\lesssim0.026$ ($R_0\lesssim6\,\mu$m) is decreasing as they
approach the wall, while for larger bubbles it increases. A similar
conclusion follows from comparing Figs~\ref{bifdiagS0} and
\ref{bifdiagS5} regarding the long-period oscillations however with
one drastic difference. Near-wall bubbles tend to oscillate with
various frequencies that depend on their size: the larger the size of
a bubble, the longer period of oscillations it has. For example,
$3T$-periodic oscillations are most profound near the second
subharmonic resonance (stars in Fig.~\ref{bifdiagS5}). They are
replaced with $4T$- (squares) and $5T$- (diamonds) periodic
oscillations as $\Omega$ increases, but such long-period oscillations
are not expected to be seen for forcing amplitudes commonly used in
medical practice. As noted earlier, the tendency of a near-wall bubble
to undergo longer period oscillations is apparently linked to the
presence of a bubble image in the model that effectively increases the
bubble inertia; note that the $\S$ term plays the role of an
additional mass in the left-hand side of equation (\ref{KMPEnd}).

It is also instructive to note that irregular oscillations can be
observed at the values of $\Me$ smaller than those shown by squares
and diamonds in Fig.~\ref{bifdiagS5}) for some initial
conditions. They could be mistaken for ``chaotic bubble
behaviour''. In reality they appear to be just transient solutions
that eventually settle to simple $T$-periodic oscillations. However
this may take a very long time (up to 20,000 of forcing periods in our
computations). The fact that they eventually decay while regular
multi-period oscillations persist confirms once again that the
appearance of the latter is due to subharmonic  resonances which
facilitate absorption of ultrasound energy by the resonating bubbles.

\section{Conclusions}

The behaviour of a microbubble approaching a solid wall and
subjected  to ultrasound forcing has been investigated using a modified
Keller-Miksis-Parlitz equation \cite{Keller1980,Parlitz1990} as a base
model. Floquet analysis has been applied to investigate the
stability of the small amplitude solutions and predict the
regions of existence of various fully nonlinear large amplitude
oscillations. It has been shown that microbubble response to acoustic
forcing can consist of $T$-, $2T$- and $3T$-periodic oscillations ($T$
is the period of ultrasound forcing) when the bubble is away from the
wall and $T$-, $2T$-, $4T$- and $5T$-periodic oscillations when the
bubble is near it. We also showed that a bubble starts ``feeling'' the
presence of the wall and changes its acoustic response when it as far
from the wall as 25-50 radial distances away. It is found that in all
cases $2T$ oscillations appear as a result of supercritical period
doubling bifurcations of $T$-periodic solutions, while longer-periodic
oscillations are completely disjoint from the $T$- and $2T$-periodic
solutions and usually require a pressure impulse to be initiated. 

Although several physical details were neglected for simplicity, such as the 
presence of a shell encapsulating medical microbubbles, the Bjerknes
force that could cause a bubble on a wall to deform, and compliance of
the wall material, the existence of clear differences in the
dynamical-systems behaviour owing to the presence of a wall suggests
that more detailed studies are warranted. The results of future
investigations could be used to estimate both the size of the bubble
and its proximity to the wall based on the spectrum of the bubble's
acoustic signature.

The demonstrated existence of multiple oscillatory solutions also
strongly indicates that bifurcation diagrams (Poincar\'e maps) that
are frequently used in studies of bubble dynamics could be misleading
unless initial conditions used to obtain such diagrams are clearly
stated and added to the set of the governing parameters characterising
the problem.

\section{Appendix. Numerical aspects\label{appendix}}
The major value of Floquet analysis is in its ability to predict very
accurately various period-changing bifurcations. However this requires
the explicit knowledge of the periodic solution $(r_0(\tau),\dot r_0(\tau))$
stability of which it aims to investigate. One may try to use a
``brutal force'' approach by integrating the governing equation
(\ref{KMPEnd}) starting with some arbitrary initial conditions over a
sufficiently long time to allow all transients to decay and a periodic
(limit cycle) solution to establish. However this can only work if
this solution is stable.

Yet to determine the bifurcation point iteratively one needs
to compute Floquet multipliers and thus know the periodic solution
$(r_0(\tau),\dot r_0(\tau))$ in parametric regions where this solution
becomes unstable and thus cannot be obtained using forward time
integration. Therefore instead of solving (\ref{KMPEnd}) as an initial
value problem one needs to view it as a periodic boundary problem with
a specified period (the forcing period or its integer
multiple). Solving such a boundary value problem is done iteratively
and the convergence of iterations depends sensitively on the initial
guess. For $T$-periodic oscillations it is readily available from a
large time limit of (\ref{smas}) and the iterative determination of
one period-doubling bifurcation point takes just a few seconds of CPU
time.

However this analytic solution cannot be used as an initial guess for
large amplitude long period oscillations. In fact, it is not even
known whether such solutions exist as they are fully disjoint from the
$T$- and $2T$-periodic solutions naturally linked to (\ref{smas}).
Therefore first we used the analysis of the phase $\theta$ of Floquet
multipliers (see Figs~\ref{egvR10_5S0T3} and \ref{egvR10_5S5T4})
of the $T$-periodic solutions to establish parametric ranges where
fully nonlinear large amplitude long period solutions can possibly
exist. Then for a so selected set of governing parameters we tested a
number of empirically chosen ``pressure impulse'' initial conditions
to obtain a stable large amplitude solution using forward time
integration. Numerically, this was done by specifying a sufficiently
large (typically of order $10^{-1}$ or even 1) negative value of $\dot
r_0$. 

Once the first long term $nT$-periodic ($n>2$) solution was obtained
in such a way, similarly to (\ref{smas}) offering a suitable initial
guess for solving a periodic boundary value problem for small
amplitude solutions, it can be used as a natural initial guess for a periodic
boundary value problem for large amplitude oscillations. However we
found that solving it for $nT$-periodic oscillations still encountered
significant numerical difficulties due to the cusp-like singularity
developing near the minima of the bubble contraction curves (see
Fig~\ref{M1825S00}). This singularity frequently results in the
divergence of iterations due to the local loss of numerical
approximation near the cusp. Since the location of such a cusp changes
during iterations, adaptively increasing the local grid density near it
to recover the approximation accuracy becomes a technical
challenge. For this reason in the current study we chose not to solve
a periodic boundary value problem for long period nonlinear
oscillations. Instead we used a more robust forward time integration
e.g.~implemented in Matlab's function {\tt ode15s} for numerically stiff
problems for all large amplitude oscillations since it has a built-in
automatic algorithm for reducing a computational step when the time
derivative becomes large.

As discussed above, only stable oscillations can be obtained using
this approach so that an automatic iterative search of the transition
point based on Floquet multipliers that was successfully implemented
for small amplitude oscillations  cannot be guaranteed to work for
large amplitude $nT$-periodic solutions. Therefore instead, once the
first of such solutions was obtained as discussed above, the governing
physical parameters were gradually changed and the final values of
$(r_0,\dot r_0)$ from a previous forward time integration run were
used as initial values for a new run in order to trace the variation of large
amplitude solutions with $\Me$, $\S$ and $\Omega$. The length of each
such run was between 900 and 4000 forcing periods (between 50 and 300
relaxation time units $\tau_0$)  to allow for statistically steady
oscillations to establish. The longest runs took up to 4 min on a
standard desktop computer.

The lines representing $nT$-periodic solutions in
Figs~\ref{bifdiagS0}, \ref{SR10_5} and \ref{bifdiagS5} were obtained
in this way and thus they correspond to parametric values at which the
respective long-period numerical solutions cannot be found any more by
forward time integration using the above parametric continuation
procedure towards smaller values of $\Me$. 

\bibliographystyle{plain}
\bibliography{bubbleref}
\end{document}